\documentclass[rnote]{aa} 
\usepackage{graphicx}
\begin{document}
  \title{A study of NIR atmospheric properties at Paranal Observatory}

  \author{G. Lombardi \inst{1}, E. Mason \inst{2}, C. Lidman \inst{3}, A.O. Jaunsen \inst{4},
  \and A. Smette \inst{1}}

	\offprints{G. Lombardi}

	\institute {European Southern Observatory, Casilla 19001, Santiago 19, Chile\\
	 \email {glombard@eso.org}
	 \and
	 ESA-STScI, 3700 San Martin Drive, MD 21218, Baltimore, USA
	 \and
	 Australian Astronomical Observatory, PO Box 296, Epping NSW 1710, Australia
	 \and
	 Institute of Theoretical Astrophysics, University of Oslo, PO Box 1029 Blindern, N-0315 Oslo, Norway}

  \date{Received 23 December 2010; accepted 29 January 2011}

	\abstract
   {}
   { In order to maximize the scientific return of the telescopes
     located at the Paranal Observatory, we 
     analyse the properties of the atmosphere above Paranal in the near-infrared (NIR).}
   { We estimate atmospheric extinction in the
     spectral range 1.10-2.30 $\mu$m ($J$, $J_{s}$, $H$, and $K_{s}$) using standard stars
     that were observed during photometric and clear nights
     with ISAAC on the Very Large Telescope UT1 between 2000 and 2004. We have
     built a database consisting of hundreds of observations, which allows us
     to examine how extinction varies with airmass and the column density of water vapour.
     In addition, we use theoretical models of the atmosphere
     to estimate Rayleigh scattering and molecular absorption
     in order to quantify their fractional contribution to the total
     extinction in each filter. Finally, we have observed 8 bright red standard stars
     to evaluate filter color terms.}
   {We find that extinction coefficients are $< 0.1$ mag
     airmass$^{-1}$ in all the considered bands. The extinction coefficient in the
     $J$-band strongly depends on the column density of water vapour. Molecular
     absorption dominates the extinction in $J$, $H$ and $K_{s}$,
     while Rayleigh scattering contributes most to the extinction in
     $J_{s}$. We have found
     negligible color terms for $J$, $H$ and $K_{s}$ and a
     non-negligible color term for $J_{s}$.}
   {}

  \keywords{Atmospheric Effects, Site Testing}

  \authorrunning{Lombardi et al.}
  \titlerunning{NIR atmospheric properties at Paranal}
  \maketitle

\section{Introduction}

    The ESO Paranal Observatory is located on the edge of the Atacama
    Coast in Northern Chile, 120 km south of the city of Antofagasta,
    at an altitude of 2635 meters above sea level.  The conditions at the observatory are
    characterised by a high fraction of clear nights 
    and low levels of humidity (Lombardi et al. \cite{lombardi09}), which
    are important for near-infrared (NIR)
    observations. However, there has been little direct measurement of the 
    properties of the atmosphere 
    above Paranal in the NIR. In order to
    fill this gap, we have calculated the extinction coefficients
    using photometric standard stars observed with the Infrared
    Spectrometer And Array Camera (ISAAC) at the Very Large Telescope
    (VLT) Unit Telescope 1 (UT1) as part of the ISAAC calibration
    plan. In addition, we have estimated the fractional contribution
    of Rayleigh scattering, molecular absorption and aerosol
    scattering to the extinction. This estimation required the
    application of a theoretical approach retrieved from the
    literature. Finally, to complete the analysis, we also calculated
    the instrumental color terms using 8 red stars observed under
    photometric conditions.

     \begin{table}[b]
     \begin{center}
       \caption[]{Properties of ISAAC SW imaging filters used in this study.}\scriptsize
         \label{isaacirfilt}
         \begin{tabular}{l c c c}
           \hline
            \noalign{\smallskip}
            Filter & Central Wavelength [$\mu$m] & Width [$\mu$m] & Width [\%]\\
            \noalign{\smallskip}
            \hline
            \noalign{\smallskip}
                  $J$ & 1.25 & 0.29 & 23\\
              $J_{s}$ & 1.24 & 0.16 & 13\\
			$H$	& 1.65 & 0.30 & 18\\
			$K_{s}$	& 2.16 & 0.27 & 13\\
            \noalign{\smallskip}
            \hline
         \end{tabular}
         \end{center}
         \begin{tiny}
         \tiny{\textbf{NOTES.} See http://www.eso.org/sci/facilities/paranal/instruments/isaac/doc/}
         \end{tiny}
\end{table}

\begin{figure*}[ht!]
	\centering
\resizebox{0.7\hsize}{!}{\includegraphics{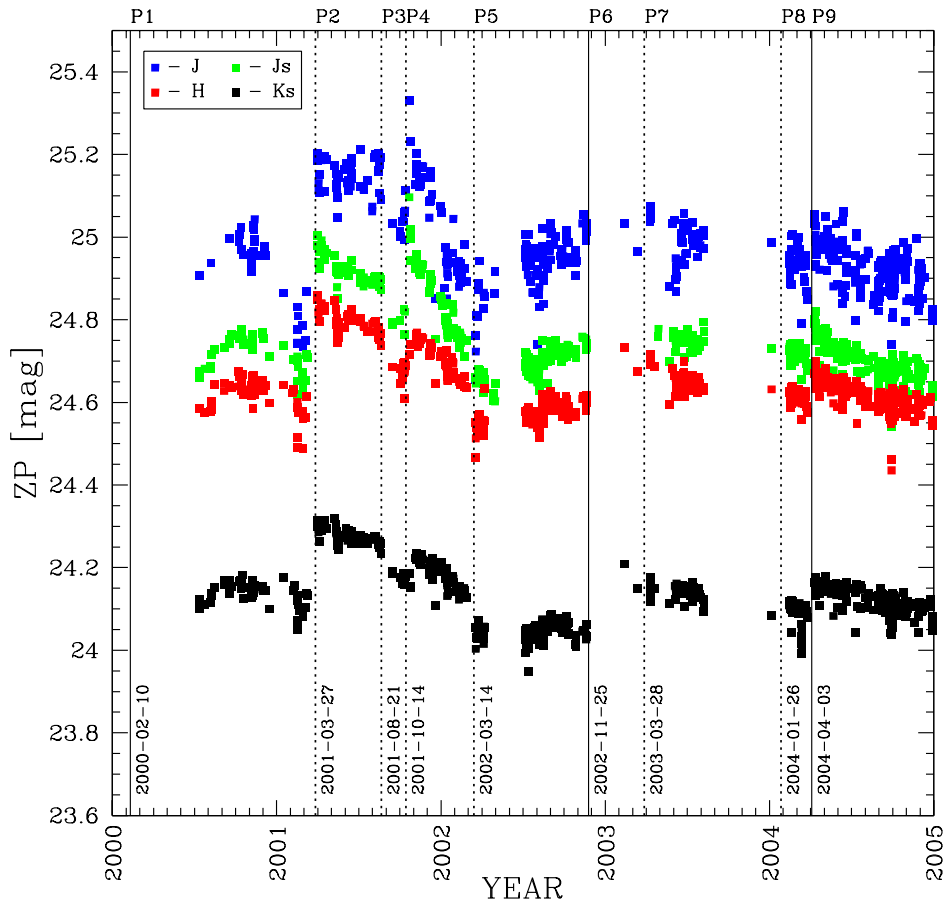}}
	\caption[Evolution in time of the zeropoints for $J$, $J_{s}$, $H$ and $K_{s}$. The solid vertical lines indicate M1 recoating events, while dotted vertical lines indicate ISAAC interventions.]{Evolution in time of the zeropoints for $J$, $J_{s}$, $H$ and $K_{s}$. The solid vertical lines indicate M1 recoating events, while dotted vertical lines indicate ISAAC interventions.}
	\label{zpvstime}
\end{figure*}
\section{Database and data reduction}

Our dataset consists of standard star observations covering a period
of 5 years (from March 2000 through to December 2004). The
observations were made during clear nights with the ISAAC
short-wavelength (SW) arm in $J$, $J_{s}$, $H$, and $K_{s}$. All data,
including the relevant calibration frames, were retrieved from the ESO
archive. Clear nights were identified by inspecting the nightly
observatory weather reports. All stars have magnitudes on the LCO
system (Persson et al. \cite{persson98}).

Table \ref{isaacirfilt} lists the characteristics of ISAAC
filters. (ISAAC User Manual \cite{isaac09}), while in Table
\ref{stars}, at the end of this Research Note, we note all the
standard stars used in the analysis. The data reduction has been
performed using the ISAAC Pipeline. Each frame has been corrected for
electronic artifacts, dark subtracted and flat fielded.  Each standard
star is imaged over a grid of five positions, one just above the
center of the array and one in each quadrant. The pipeline computes a
set of instrumental magnitudes which are averaged to deliver the 
zeropoint ($ZP$) uncorrected for extinction.

$ZP$ uncertainties are computed as $\sigma =
\sigma_{ZP_{i}}/\sqrt{n}$, where $n$ is the number of times, usually
five, a single standard is observed, and $\sigma_{ZP_{i}}$ is the
scatter about the mean ZP reported by the pipeline.
In our analysis, we have rejected points having $\sigma>0.050$ mag. Our
final sample contains 575 data points in $J$, 603 data points in $J_{s}$,
604 data points in $H$ and 667 data points in $K_{s}$.

\section{Data analysis}\label{data_analysis}
\subsection{Time evolution of the zeropoint}\label{timecorrection}

In Figure \ref{zpvstime}, the zeropoints of the four bands are shown
as four time series. Different colors represent
different bands. We clearly see that, for each band, the zeropoints are
characterized by trends within well defined time intervals. A further
inspection demonstrated that the intervals are delimited by technical
or maintenance interventions, either on the instrument or the telescope
(e.g. M1 recoating). This means that, in the mentioned intervals (or
periods, P), the computed zeropoints are affected by deterioration of
the telescope optics after a recoating of UT1 primary mirror (M1) or
instruments troubles (ISAAC technical interventions). A list of the
events that occurred between 2000 and 2005 is reported in Table
\ref{events}.

\begin{table}[b]
       \caption[Technical events occurred between 2000 and 2005.]{Technical events occurred between 2000 and 2005.}
    \label{events}
 \begin{center}\scriptsize
         \begin{tabular}{c l l}
	\hline
            Item & Date & Event\\
\hline
P1 & 2000 02 10 & M1 recoating\\ 
P2 & 2001 03 27 & ISAAC intervention\\ 
P3 & 2001 08 21 & ISAAC intervention\\ 
P4 & 2001 10 14 & ISAAC intervention\\ 
P5 & 2002 03 14 & ISAAC intervention\\ 
P6 & 2002 11 25 & M1 recoating\\ 
P7 & 2003 03 28 & ISAAC intervention\\ 
P8 & 2004 01 26 & ISAAC intervention\\ 
P9 & 2004 04 03 & M1 recoating\\ 
\hline
         \end{tabular}
         \end{center}
\end{table}
\begin{figure}
\centering
\includegraphics[width=8cm]{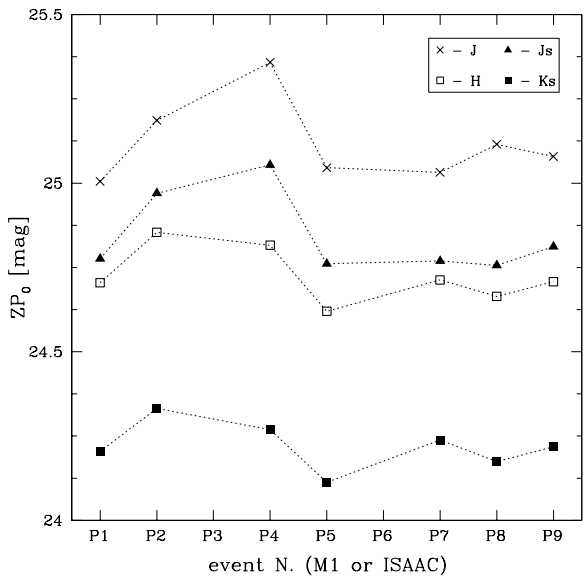}
   \caption[$ZP_0$ for each period.]{$ZP_0$ for each period.}
   \label{zpcheck}
\end{figure}

The deterioration in the aluminum coating of the telescope mirrors is mainly due to dust and oxidation. It results in a
progressive reduction in the mirror reflectivity, an increase in the
thermal background emission (Frogel \cite{frogel98}) and a consequent
decrease in the zeropoints. A change (increase or decrease) in the zeropoint
can also be due to interventions on ISAAC, as they affect the
instrument configuration. Our idea was to consider each
period separately and to remove, from each one of them, the time dependency by
subtracting a linear fit of the zeropoint with time.
In practice we adjust all zeropoints to the value of the
intercept of the fit at the beginning of the considered period. During
the analysis we have confirmed that a linear fit was
accurate enough, therefore higher polynomials have not been used.
The fits for periods P3 and P6 are not well constrained,
because the small number of points within these two periods.
Therefore, they are not been considered further in this
analysis.

The evolution in the zeropoint within each period is fitted with
\begin{equation}\label{zp2}
	ZP(t) = Ct + ZP_0
\end{equation}
where $t$ is the time, $ZP_0$ corresponds to the zeropoint at the
beginning ($t=0$) of the considered period, and $C$ the slope of the
fit. The differences $ZP(t=0)-ZP_i$ have been computed and added to
each observed $ZP_i$.

Finally, in each period, we subtracted the intercept of the fit at
$t=0$ from the corrected $ZP_i$. In this way we have eliminated 
both the offset between periods and the evolution with time within a period.
This represents a clear advantage. We can now work
with a single dataset, rather than 9.

It is interesting to check the variation in $ZP_0$ after each
event. If the event did not affect the instrument
performance, $ZP_0$ values should be all identical. As shown in
Figure \ref{zpcheck}, we observe significant offsets between events.

\begin{figure}
   \centering
   \includegraphics[width=8cm]{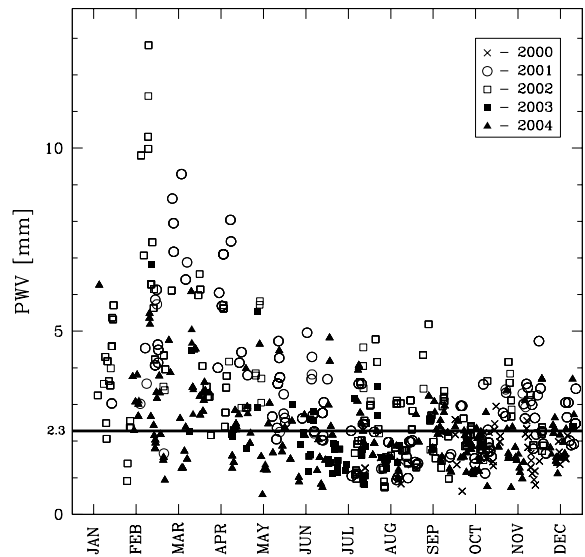}
   \caption[Monthly trend of the PWV on Paranal in photometric and clear nights.]{Monthly trend of the PWV on Paranal in photometric and clear nights.}
              \label{pwvvstime}
    \end{figure}

\subsection{Effects of the Precipitable Water Vapour}\label{pwvcorrection}

The transmission of the atmosphere in the NIR is greatly affected by
the presence of water. Considering the classical Johnson filters,
an increase in the column density of water vapour causes $(J - K)$ and $(H - K)$ to be bluer,
because the effect of water vapour is stronger in $K$
($\Delta\lambda = 0.41$ $\mu$m) than in $J$ and $H$ (Frogel
\cite{frogel98}; see also Figure 1 in Manduca\&Bell
\cite{manduca79}). On the other hand, the ISAAC $K_{s}$ filter has
a narrower bandwidth ($\Delta\lambda = 0.27$ $\mu$m, see Table
\ref{isaacirfilt}) and avoids the strongest water lines. Therefore we
expect that, in our case, $J$ and $H$ will suffer more from large
amounts of water in the atmosphere.

The amount of water above Paranal has been continuously monitored
since July 2000 using the images from the Geostationary Operational Environmental
Satellite (GOES). The sampling is every 3 hours starting at 00h UT.

To retrieve the Precipitable Water Vapour (PWV) from GOES images
a model exists developed by A. Erasmus under contract
with ESO (Erasmus\&Peterson \cite{erasmus97}; Erasmus\&Sarazin \cite{erasmus00}; Erasmus\&Sarazin \cite{erasmus02}). The method is based on combining satellite current image in the 6.7 $\mu$m and 10.7 $\mu$m channel
with wind and temperature profiles forecasted by a prediction center.

Typical satellite observations at about 6.5 $\mu$m are sensitive to emissions from water vapour resident in the layer between about 600 mbar ($\sim$4400 m) and 300 mbar ($\sim$9000 m). For what concerns GOES vertical resolution, the 6.7 $\mu$m channel
is located near the center of a strong water vapor absorption band
and under clear sky conditions it is primarily sensitive to the relative
humidity averaged over a depth of atmosphere extending from 200 to 500 mbar (Soden\&Bretherton \cite{soden93}).
The horizontal resolution is sufficient since the atmosphere
surrounding Paranal is not expected to vary significantly over an area that extends several 10's of km.

Figure \ref{pwvvstime} shows
the yearly trend in the amount of PWV sampled at Paranal during photometric and
clear nights. We do see a yearly periodic modulation with very high
values (up to 13\,mm in 2002) in the trimester January-March
corresponding to the so called \textit{Bolivian winter}. As shown in
the figure, the median PWV at Paranal during photometric and
clear nights is 2.3\,mm.

\begin{figure}
\centering
\includegraphics[width=8cm]{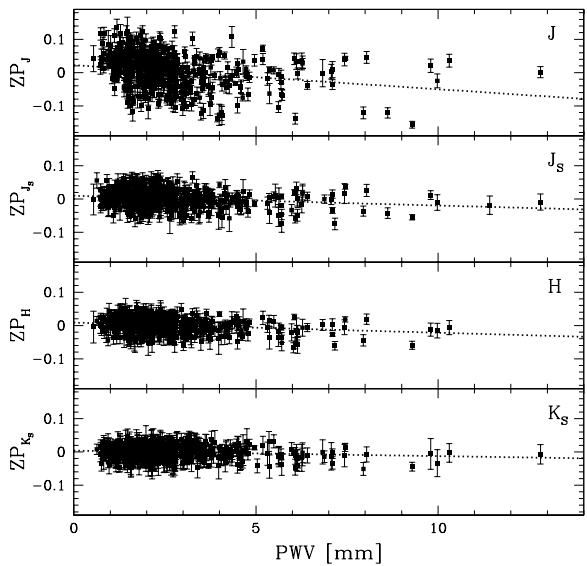}
   \caption{Weighted linear fit of the zeropoints as a function of the PWV in the four bands.}
   \label{pwvzp}
\end{figure}

Figure \ref{pwvzp} shows the relationship between zeropoints and
the PWV in the considered bands. For each band we have calculated a
weighted linear fit. The fits are repeated twice more after rejecting 3-$\sigma$
outliers.
In Table \ref{pwvtab} we report the slopes and the RMS of the fits.
For $J_{s}$, $H$, and $K_{s}$ the trend with the column density of PWV is
quite slight, while for $J$ it is very significant. The scatter about the best
fit for $J$ is between 2 and 4 times larger that the scattered measured for the other filters.

\begin{table}
       \caption[]{Zeropoints-versus-PWV slope and dispersion (RMS).}
      \begin{center}\scriptsize
   \label{pwvtab}
         \begin{tabular}{c | c c }
	 \hline
					Band & Slope & Dispersion\\
					& [mag per unit of PWV] & [mag]\\
\hline
    $J$ & $-0.007 \pm 0.001$ & 0.048\\
$J_{s}$ & $-0.003 \pm 0.001$ & 0.024\\
    $H$ & $-0.003 \pm 0.001$ & 0.019\\
$K_{s}$ & $-0.002 \pm 0.001$ & 0.011\\
\hline
         \end{tabular}
         \end{center}
\end{table}

\begin{table*}
  \caption[]{Calculated $\kappa$ in [mag airmass$^{-1}$] for $J$, $J_{s}$, $H$ and $K_{s}$ at Paranal.}
      \begin{center}\footnotesize
   \label{extcoeff}
         \begin{tabular}{ c | c c c c | c c c c | c c c c | c c c }
	 \hline
			Band		& & \multicolumn{2}{c}{\textbf{PWV-corrected ZP}} & & & \multicolumn{2}{c}{\textbf{PWV = 0--2 mm}} & & & \multicolumn{2}{c}{\textbf{PWV = 2--4 mm}} & & & \multicolumn{2}{c}{\textbf{PWV = 4--7 mm}}\\
					& & $\kappa$ & disp. & & & $\kappa$ & disp. & & & $\kappa$ & disp. & & & $\kappa$ & disp.\\
\hline
    $J$ & & \textbf{0.072} & 0.040 & & & \textbf{0.038} & 0.038 & & & \textbf{0.060} & 0.036 & & & \textbf{0.090} & 0.033\\
$J_{s}$ & & \textbf{0.048} & 0.019 & & & \textbf{0.034} & 0.019 & & & \textbf{0.040} & 0.020 & & & \textbf{0.058} & 0.018\\
    $H$ & & \textbf{0.034} & 0.015 & & & \textbf{0.035} & 0.015 & & & \textbf{0.030} & 0.015 & & & \textbf{0.053} & 0.016\\
$K_{s}$ & & \textbf{0.043} & 0.013 & & & \textbf{0.040} & 0.015 & & & \textbf{0.046} & 0.014 & & & \textbf{0.042} & 0.012\\
\hline
         \end{tabular}
         \end{center}
\end{table*}

\section{Determination of the atmospheric extinction coefficients}\label{empirical}
The relationship between $ZP$ and airmass ($X$) defines the extinction curve (Bouguer curve). The extinction coefficient ($\kappa$) is
the slope of a linear fit to this curve. The fit is done three times. Between each fit, 3-$\sigma$ outliers are rejected.

The extinction coefficients have been calculated in two different ways:

\begin{itemize}

\item[{\it a})] correcting the zeropoints for the amount of PWV before computing $\kappa$;

\item[{\it b})] considering separately the zeropoints in different PWV ranges in order to evaluate $\kappa$ as a function of the PWV.
\end{itemize}

\begin{figure}
   \centering
   \includegraphics[width=8cm]{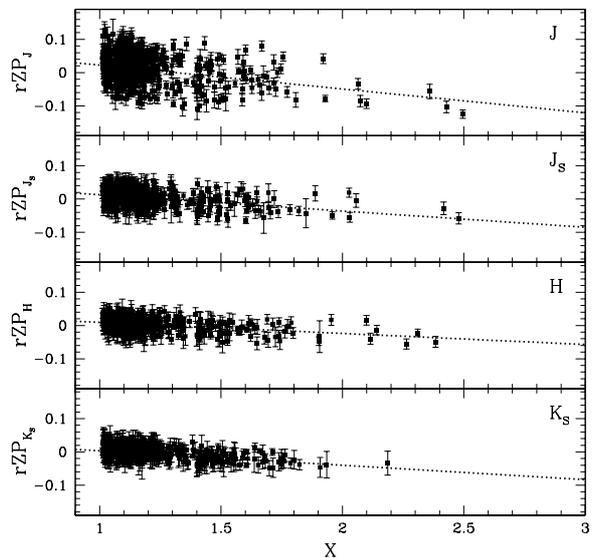}
   \caption{Bouguer
 curves computed for the PWV rescaled zeropoints in $J$, $J_{s}$, $H$ and $K_{s}$.}
   \label{rzpx}
\end{figure}

In case {\it a} we offset the zeropoints 
using the fits calculated in \S
\ref{pwvcorrection} and reported in Figure \ref{pwvzp} and Table
\ref{pwvtab}. 
The offset to each zeropoint is 
\begin{equation}
\Delta ZP_i = ZP_i(PWV=2.3) - ZP_i(PWV)
\end{equation}
where the index $i$ refers to one of the four filters. The zeropoint is corrected to the median amount of PWV. The PWV-corrected
zeropoint, $rZP_i$, is then

\begin{equation}
	rZP_i = ZP_i + \Delta ZP_i
\end{equation}
In Figure \ref{rzpx} we have plotted the weighted linear fits to the
 curves for the $rZP$ in the four bands.  The extinction coefficients
are reported in column 2 of Table \ref{extcoeff}. The
dispersion (RMS) of the points around the best fits are also reported.

\begin{figure}
   \centering
   \includegraphics[width=8cm]{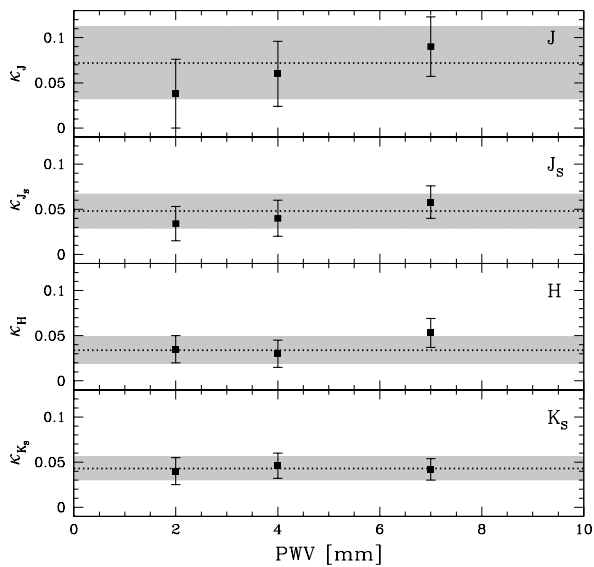}
   \caption{Extinction coefficients in $J$, $J_{s}$, $H$ and $K_{s}$ calculated for PWV = 0--2 mm, PWV = 2--4 mm, PWV = 4--7 mm. The dotted line represents the $\kappa$ calculated with the PWV rescaled $ZP$, while the gray zone is the dispersion (see column 2 of Table \ref{extcoeff}).}
              \label{extvspwv}
    \end{figure}

In case {\it b} we have calculated the extinction coefficients for
three different PWV ranges: PWV =  0--2 mm, PWV = 2--4 mm,
and PWV = 4--7 mm. 
The results are 
reported in columns 3, 4, and 5 of Table \ref{extcoeff} and
plotted in Figure \ref{extvspwv}. As expected, the $J$-band extinction
coefficient has the greatest sensitivity to the amount of PWV. It
increases by $\sim$$0.05$ mag airmass$^{-1}$ over the range of PWVs
considered here. The increase in the extinction coefficients for
$J_{s}$ and $H$ are less ($\sim$$0.03$ and $\sim$$0.02$ mag
airmass$^{-1}$, respectively), while there is a negligible effect in
$K_{s}$.

The dependence of the extinction coefficients 
on the amount of PWV underlines the importance of observing IR standards (more
important for $J$ and less important for $K_{s}$) at roughly the same
airmass and at roughly the same time as the science target.

\section{The fractional contribution to the extinction}

According to Hayes\&Latham (\cite{hayes75}) the largest
contributors to atmospheric extinction are Rayleigh scattering
($\kappa_{Ray}$), molecular absorption ($\kappa_{mol}$) and aerosol
scattering ($\kappa_{aer}$). The efficiency of aerosol scattering
from particles that are smaller than a few microns at $\lambda > 1.0$
$\mu$m is negligible (Lombardi et al. \cite{lombardi08}), therefore
for the ISAAC filters (Table \ref{isaacirfilt}) we have computed
the vertical atmospheric extinction coefficients as the sum of
$\kappa_{Ray}$ and $\kappa_{mol}$

\begin{equation}\label{k}
\kappa_{\lambda}=\kappa_{Ray,\lambda}+\kappa_{mol,\lambda}
\end{equation}
Variations in $\kappa_{Ray}$ and $\kappa_{mol}$ with time are linked
to occasional and periodic climatic changes in the atmosphere above
the site (Lombardi et al. \cite{lombardi09}). For this reason,
theoretical models are valid only to calculate average contributions
to the extinction coefficients.

\subsection{Rayleigh scattering}\label{rayleigh}

Rayleigh scattering of unpolarised light is due to particles with
dimensions that are $\ll\lambda$. It has been extensively discussed by
Penndorf (\cite{penndorf57}). According to Hayes\&Latham
(\cite{hayes75}), Rufener (\cite{rufener86}) and Burki et
al. (\cite{burki95}) the vertical Rayleigh extinction can be expressed
as
\begin{equation}\label{k_Ray}
\kappa_{Ray,\lambda}=B\lambda^{-4}
\end{equation}
where the Rayleigh scattering coefficient $B$ is a complex function of
the refractive index, $n(T,P,\lambda)$, and the number of particles,
$N(T,P)$. $T$ and $P$ are the site's
mean temperature and pressure. $B$ can be written as $B(T,P,\lambda)$
and expressions for $n$, $N$ and $B$ are given by Penndorf
(\cite{penndorf57}).

In the years between 2000 and 2004 a mean temperature of
$(12.8\pm0.5)$$^{\circ}$C and a mean pressure of $(743.5\pm0.2)$ hPa have
been measured at Paranal (Lombardi et al. \cite{lombardi09}). We have
used the expressions in Penndorf (\cite{penndorf57}) and
Hayes\&Latham (\cite{hayes75}) to compute $B(T,P,\lambda)$ and $\kappa_{Ray,\lambda}$ 
for the four ISAAC filters. The values are reported in Table \ref{kRay}.

As expected, Rayleigh scattering is more efficient in $J$ and $J_{s}$.
It is an order of magnitude less efficient in $H$ and $K_{s}$.

\begin{table}[t]
     \begin{center}\footnotesize
       \caption[]{Rayleigh scattering coefficients and vertical Rayleigh extinction in [mag airmass$^{-1}$] calculated for Paranal.}\scriptsize
    \label{kRay}
         \begin{tabular}{c c c}
	 \hline
          \noalign{\smallskip}
					& $B(T,P,\lambda)$ & $\kappa_{Ray,\lambda}$\\
          \noalign{\smallskip}
\hline
     $J$ & $0.106 \pm 0.011$ & $0.043 \pm 0.005$\\
 $J_{s}$ & $0.109 \pm 0.011$ & $0.046 \pm 0.005$\\
     $H$ & $0.035 \pm 0.004$ & $0.005 \pm 0.001$\\
 $K_{s}$ & $0.012 \pm 0.001$ & $0.001 \pm 0.001$\\
\hline
         \end{tabular}
         \end{center}
\end{table}

\subsection{Molecular absorption}\label{molecular}

The contribution to the extinction originating from molecular
absorption was modeled using an IDL driver to the Reference Forward
Model (RFM).  RFM is a GENLN2-based line-by-line radiative transfer
code developed by Anu Dudhia at the Atmospheric,
Oceanic and Planetary Physics Institute at Oxford University (UK) to
analyse data from MIPAS on-board
ENVISAT\footnote{http://www.atm.ox.ax.uk/RFM}.  The code was run using
the 2008 version of the HITRAN database (Rothman et
al. \cite{hitran09}).

For the current work, only lines caused by H$_{2}$O, CO$_{2}$,
N$_{2}$O, CH$_{4}$ and O$_{2}$ were considered.  We slightly modified the
tropical atmospheric profile available with the RFM code to carry-out
simulations corresponding to a temperature of 12.8$^{\circ}$C and an atmospheric
pressure of 743.5\,hPa (Lombardi et al. \cite{lombardi09}). The amount
of water vapour was scaled to match the median PWV
for Paranal (2.3\,mm). The code also allows one to vary the airmass
of the line-of-sight or to select a specific molecule.  The extinction
in each filter was estimated by multiplying the atmospheric
transmission produced by the model with the filter transmission curve (see ISAAC User Manual \cite{isaac09}). Results are reported in Table \ref{kmol}.

\begin{table}[t]
     \begin{center}\footnotesize
       \caption[]{Calculated vertical molecular extinction in [mag airmass$^{-1}$] for Paranal.}\scriptsize
    \label{kmol}
         \begin{tabular}{c c c c c}
	 \hline
          \noalign{\smallskip}
	Molecule & $J$ & $J_{s}$ & $H$ & $K_{s}$\\
          \noalign{\smallskip}
\hline
      CH$_{4}$ & 0.001 & 0.000 & 0.004 & 0.015\\
      CO$_{2}$ & 0.000 & 0.001 & 0.005 & 0.040\\
      H$_{2}$O & 0.189 & 0.018 & 0.027 & 0.004\\
      N$_{2}$O & 0.000 & 0.000 & 0.000 & 0.001\\
      O$_{2}$  & 0.004 & 0.007 & 0.000 & 0.000\\
\hline
$\kappa_{mol,\lambda}$ & 0.194 & 0.026 & 0.036 & 0.060\\
\hline
         \end{tabular}
         \end{center}
\end{table}

\begin{table}[t]
     \begin{center}\footnotesize
       \caption[]{Vertical atmospheric extinction coefficients [in mag
         airmass$^{-1}$] and fractional contributions of Rayleigh
         scattering and molecular absorption.}\scriptsize
    \label{fractions}
         \begin{tabular}{c c c c}
	 \hline
          \noalign{\smallskip}
	& $\kappa_{\lambda}$ & fraction of $\kappa_{Ray,\lambda}$ & fraction of $\kappa_{mol,\lambda}$\\
          \noalign{\smallskip}
\hline
     $J$ & $0.237 \pm 0.010$ & 18\% & 82\% \\
 $J_{s}$ & $0.072 \pm 0.010$ & 64\% & 36\% \\
     $H$ & $0.041 \pm 0.010$ & 12\% & 88\% \\
 $K_{s}$ & $0.061 \pm 0.010$ & 2\% & 98\% \\
\hline
         \end{tabular}
         \end{center}
\end{table}
\begin{table*}[t!]
     \begin{center}
       \caption[]{Observed Persson's red standard stars for the determination of the color terms (the computed $C_{m}$ is also reported).}\scriptsize
    \label{ctmagnitudes}
         \begin{tabular}{c  c  c  c  c  c  c  c}
\hline
          \noalign{\smallskip}
           & R.A. & Decl. & & & & &\\
            Star & (J2000) & (J2000) & $J-K_{s}$ & $J-J_{IS}$ & $J-J_{S,IS}$ & $H-H_{IS}$ & $K_{s}-K_{S,IS}$\\
          \noalign{\smallskip}
\hline
cskd-21  & 12 32 10.9 & $-63$ 43 16 & 0.724 & $-0.262 \pm 0.022$ & $ 0.018 \pm 0.022$ & $0.192 \pm 0.024$ & $0.722 \pm 0.032$\\
cskf-12  & 12 31 30.1 & $-63$ 51 03 & 1.110 & $-0.281 \pm 0.017$ & $-0.011 \pm 0.015$ & $0.204 \pm 0.011$ & $0.731 \pm 0.012$\\
cskf-14a & 12 31 45.9 & $-63$ 49 36 & 1.734 & $-0.281 \pm 0.023$ & $-0.005 \pm 0.019$ & $0.207 \pm 0.018$ & $0.708 \pm 0.022$\\
cskd-20  & 12 32 04.0 & $-63$ 43 46 & 1.826 & $-0.284 \pm 0.026$ & $ 0.025 \pm 0.026$ & $0.219 \pm 0.017$ & $0.725 \pm 0.020$\\
cskd-9   & 12 31 16.7 & $-63$ 40 11 & 2.219 & $-0.243 \pm 0.039$ & $ 0.047 \pm 0.032$ & $0.187 \pm 0.030$ & $0.690 \pm 0.033$\\
cske-23  & 12 31 56.0 & $-63$ 37 43 & 2.735 & $-0.244 \pm 0.027$ & $ 0.026 \pm 0.025$ & $0.221 \pm 0.017$ & $0.736 \pm 0.020$\\
cskd-34  & 12 31 23.6 & $-63$ 46 45 & 2.826 & $-0.286 \pm 0.021$ & $ 0.034 \pm 0.022$ & $0.198 \pm 0.019$ & $0.725 \pm 0.023$\\
cskd-16  & 12 31 57.8 & $-63$ 42 21 & 2.862 & $-0.254 \pm 0.021$ & $ 0.056 \pm 0.020$ & $0.203 \pm 0.019$ & $0.732 \pm 0.021$\\
\hline
           & & & $C_{m}$ & $0.008 \pm 0.010$ & $0.020 \pm 0.009$ & $0.003 \pm 0.006$ & $0.002 \pm 0.008$\\
\hline
         \end{tabular}
         \end{center}
\end{table*}
\subsection{Fractional contributions}

In column 1 of Table \ref{fractions}, we list
$\kappa_{\lambda}$ for $J$, $J_{s}$, $H$ and $K_{s}$.
The values have been derived using equation (\ref{k}) and the computed values for $\kappa_{Ray,\lambda}$ and
$\kappa_{mol,\lambda}$. The uncertainties derive from the propagation
of the uncertainties of $T$ and $P$ in Lombardi et
al. (\cite{lombardi09}) and as to be assumed has an upper limit to the
variation of the theoretical $\kappa_{\lambda}$.

For $J_{s}$, $H$ and $K_{s}$, there is good agreement
between the theoretical $\kappa_{\lambda}$ and the
empirical $\kappa$ calculated from the PWV rescaled zeropoints
(column 2 of Table \ref{extcoeff}). For the $J$-band, there is a significant
discrepancy, which is due to the high opacity of the atmosphere at the
red end of the $J$ filter. The red end of the $J$ filter is
effectively defined by the atmosphere. Changes in the amount of PWV
in the atmosphere shifts this edge, which leads to larger photometric
uncertainties, as demonstrated by the large scatter about the best fits
in Figures~\ref{pwvzp} and \ref{rzpx}.
As noted in \S \ref{empirical}, this result demonstrates the
importance of observing IR standards at roughly the same
airmass and at roughly the same time as the science target when aiming
for precise photometry in the $J$-band.

Columns 2 and 3 of Table \ref{fractions} show the fractional
contributions of $\kappa_{Ray,\lambda}$ and $\kappa_{mol,\lambda}$ to
the total vertical atmospheric extinction. Molecular absorption
dominates the absorption in $J$ (82\%), $H$ (88\%) and $K_{s}$ (98\%). 
Molecular absorption and Rayleigh scattering contribute almost equally to the
absorption observed in  $J_{s}$.
This is due to
the design of the ISAAC $J_{s}$ filter which is narrower than $J$ and
avoids the strong atmospheric absorption lines at the end of $J$-band.

\section{Determination of the color terms}\label{colorterms}

According to Amico et al. (\cite{amico02}), the $J$, $H$ and $K_{s}$ filters of ISAAC closely
match the filters tabulated in Persson et
al. (\cite{persson98}). We therefore expect negligible color terms for
these filters. The $J_{s}$ filter of ISAAC, on the other hand, is significantly different to $J$, so
we expect a significant color term for $J_{s}$.

We observed 8 bright red stars from Persson et
al. (\cite{persson98}) with ISAAC in $J$, $J_{s}$, $H$ and $K_{s}$
under photometric conditions (see Table \ref{ctmagnitudes}).  Each
star has been observed over a grid of four positions (one for each
quadrant of the detector) with a windowed detector and short
integrations ($< 1$ s) to avoid saturation. The integrations were shorter than the 
minimum integration time required for full readout of the array, so only a subarray of
the detector was read out.\\
We 
integrated the flux in apertures of fixed size to derive instrumental magnitudes
for each star. The final instrumental magnitude associated
to the star ($m_{IS}$)  is the weighted average of
the four instrumental magnitudes corrected using the atmospheric
extinction coefficients determined in \S \ref{empirical} (see column 2 in Table
\ref{extcoeff}). 

\begin{figure}[t]
   \centering
   \includegraphics[width=8cm]{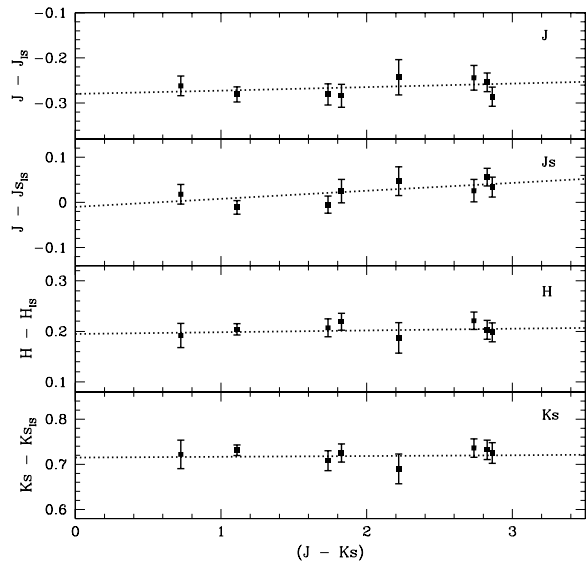}
   \caption{Plots of $(m_{cat}-m_{IS})_{i}$ versus $(J-K_{s})_{i}$. The slope of the linear fit is $C_{m}$ and is tabulated in Table \ref{ctmagnitudes}.}
              \label{ctplot}
    \end{figure}

In Figure~\ref{ctplot} we  plot $(M_{cat}-m_{IS})_{i}$  against
$(J-K_{s})_{i}$ and fit for the slope $C_{m}$. An arbitrary constant vertical
offset has been applied to the points in each of the graphs.

Table \ref{ctmagnitudes} lists $(m_{cat}-m_{IS})_{i}$ for 
$J$, $J_{s}$, $H$ and $K_{s}$ and the color
 $(J-K_{s})_{i}$. The slopes, $C_{m}$,
are also reported. 

As expected, $C_{m}$ does not differ from 0 for $J$, $H$ or $K_{s}$,
while $C_{m} = 0.020 \pm 0.009$ for $J_{s}$.

\section{Conclusions}\label{conclusions}

In this Research Note, we have characterized the extinction properties
in the NIR at the Paranal Observatory (Chile). We have used
standard stars observed with ISAAC in $J$, $J_{s}$, $H$, and $K_{s}$
during photometric and clear nights at the Very Large
Telescope UT1 between 2000 and 2004.

For each star we calculated the zeropoint uncorrected for extinction
using the ISAAC Pipeline. A
correction is then performed on the whole dataset in order to eliminate
the affects on the zeropoints due to technical
events (such as recoatings of the primary mirror and ISAAC
interventions).

We used two different methods to derive extinction co-efficients. In the
first method, we account for the varition in the ZPs with the PWV
by rescaling the zeropoints 
to the median PWV (2.3\,mm) measured at Paranal
between 2001 and 2005. We then fit a linear relation to Bouguer
 curves to
determine the extiction co-efficients. 
We have obtained $\kappa_{J}=0.072\pm 0.040$,
$\kappa_{J_{s}}=0.048\pm 0.019$, $\kappa_{H}=0.034\pm 0.015$ and
$\kappa_{K_{s}}=0.043\pm 0.013$ mag airmass$^{-1}$. In the second
method, we have calculated $\kappa$ considering the zeropoints in
three different PWV ranges: PWV = 0--2 mm, PWV = 2--4 mm,
and PWV = 4--7 mm. As expected, the extinction
coefficient in $J$ is more sensitive to the amount PWV than the extinction
coefficients of other filters, increasing by $\sim$$0.05$ mag
airmass$^{-1}$ between lower and upper ranges for the PWV. For comparison the coefficients for
$J_{s}$ and $H$ increase by $\sim$$0.03$ mag airmass$^{-1}$  and $\sim$$0.02$ mag airmass$^{-1}$, 
respectively, while there is negligible
change for $K_{s}$.

Using a theoretical approach, we have found that molecular
absorption contributes most to the total absorption in $J$ (82\%), $H$ (88\%) and $K_{s}$ (98\%),
while Rayleigh scattering contributes most to the total
absorption in the $J_{s}$-band.

We have calculated the color terms using 8 bright stars observed in
photometric conditions. We have found negligible color terms in $J$,
$H$ and $K_{s}$ and a non-negligible color term in $J_{s}$
($0.020 \pm 0.009$).

\begin{table*}[p!]
     \begin{center}
      \caption[]{Observed Persson's standard stars from the ISAAC Calibration Plan.}\scriptsize
			\label{stars}
         \begin{tabular}{c  r  r  c  c  c  c  c  c}
      \hline
          \noalign{\smallskip}
           & R.A. & Decl. & & & \\
            Star & (J2000) & (J2000) & $J$ & $\sigma_{J}$ & $H$ & $\sigma_{H}$ & $K_{s}$ & $\sigma_{K_{s}}$\\
          \noalign{\smallskip}
\hline
P525-E & 00 24 28.3 & $ 07$ 49 02 & 11.622 & 0.005 & 11.298 & 0.005 & 11.223 & 0.005\\
S294-D & 00 33 15.2 & $-39$ 24 10 & 10.932 & 0.006 & 10.657 & 0.004 & 10.594 & 0.004\\
S754-C & 01 03 15.8 & $-04$ 20 44 & 11.045 & 0.005 & 10.750 & 0.005 & 10.695 & 0.005\\
P530-D & 02 33 32.1 & $ 06$ 25 38 & 11.309 & 0.010 & 10.975 & 0.006 & 10.910 & 0.005\\
S301-D & 03 26 53.9 & $-39$ 50 38 & 12.153 & 0.007 & 11.842 & 0.005 & 11.788 & 0.006\\
P533-D & 03 41 02.4 & $ 06$ 56 13 & 11.737 & 0.009 & 11.431 & 0.006 & 11.336 & 0.005\\
S055-D & 04 18 18.9 & $-69$ 27 35 & 11.552 & 0.002 & 11.326 & 0.002 & 11.269 & 0.002\\
S361-D & 04 49 54.6 & $-35$ 11 17 & 11.246 & 0.006 & 11.031 & 0.006 & 10.980 & 0.006\\
S363-D & 05 36 44.8 & $-34$ 46 39 & 12.069 & 0.007 & 11.874 & 0.005 & 11.831 & 0.005\\
S840-F & 05 42 32.1 & $ 00$ 09 04 & 11.426 & 0.009 & 11.148 & 0.009 & 11.058 & 0.008\\
S842-E & 06 22 43.7 & $-00$ 36 30 & 11.723 & 0.011 & 11.357 & 0.009 & 11.261 & 0.010\\
S121-E & 06 29 29.4 & $-59$ 39 31 & 12.114 & 0.006 & 11.838 & 0.005 & 11.781 & 0.005\\
S255-S & 06 42 36.5 & $-45$ 09 12 & 11.719 & 0.004 & 11.434 & 0.004 & 11.372 & 0.004\\
S427-D & 06 59 45.6 & $-30$ 13 44 & 10.833 & 0.007 & 10.499 & 0.007 & 10.442 & 0.009\\
S209-D & 08 01 15.4 & $-50$ 19 33 & 10.914 & 0.007 & 10.585 & 0.006 & 10.496 & 0.009\\
S312-T & 08 25 36.1 & $-39$ 05 59 & 11.949 & 0.006 & 11.669 & 0.005 & 11.609 & 0.004\\
S495-E & 08 27 12.5 & $-25$ 08 01 & 11.521 & 0.007 & 11.048 & 0.008 & 10.960 & 0.010\\
P545-C & 08 29 25.1 & $ 05$ 56 08 & 11.881 & 0.007 & 11.624 & 0.005 & 11.596 & 0.006\\
S705-D & 08 36 12.5 & $-10$ 13 39 & 12.362 & 0.010 & 12.098 & 0.011 & 12.040 & 0.014\\
S165-E & 08 54 21.7 & $-54$ 48 08 & 12.489 & 0.008 & 12.214 & 0.008 & 12.142 & 0.011\\
S372-S & 09 15 50.5 & $-36$ 32 34 & 11.153 & 0.007 & 10.891 & 0.007 & 10.836 & 0.010\\
S708-D & 09 48 56.4 & $-10$ 30 32 & 11.081 & 0.008 & 10.775 & 0.008 & 10.718 & 0.010\\
P550-C & 10 33 51.8 & $ 04$ 49 05 & 12.344 & 0.007 & 12.121 & 0.005 & 12.081 & 0.005\\
S264-D & 10 47 24.1 & $-44$ 34 05 & 11.642 & 0.009 & 11.335 & 0.008 & 11.280 & 0.010\\
S217-D & 12 01 45.2 & $-50$ 03 10 & 11.323 & 0.007 & 11.002 & 0.005 & 10.936 & 0.004\\
S064-F & 12 03 30.2 & $-69$ 04 56 & 12.111 & 0.007 & 11.803 & 0.007 & 11.724 & 0.007\\
S860-D & 12 21 39.3 & $-00$ 07 13 & 12.213 & 0.007 & 11.917 & 0.006 & 11.865 & 0.005\\
S791-C & 13 17 29.6 & $-05$ 32 37 & 11.661 & 0.008 & 11.310 & 0.007 & 11.267 & 0.008\\
P499-E & 14 07 33.9 & $ 12$ 23 51 & 11.947 & 0.008 & 11.605 & 0.008 & 11.540 & 0.008\\
S867-V & 14 40 58.0 & $-00$ 27 47 & 12.045 & 0.008 & 11.701 & 0.005 & 11.633 & 0.005\\
S273-E & 14 56 51.9 & $-44$ 49 14 & 11.341 & 0.007 & 10.924 & 0.005 & 10.849 & 0.004\\
P565-C & 16 26 42.7 & $ 05$ 52 20 & 12.180 & 0.007 & 11.895 & 0.006 & 11.844 & 0.006\\
P330-E & 16 31 33.6 & $ 30$ 08 48 & 11.816 & 0.007 & 11.479 & 0.005 & 11.429 & 0.006\\
S279-F & 17 48 22.6 & $-45$ 25 45 & 12.477 & 0.009 & 12.118 & 0.006 & 12.031 & 0.006\\
S071-D & 18 28 08.9 & $-69$ 26 03 & 12.252 & 0.006 & 11.916 & 0.007 & 11.839 & 0.007\\
S808-C & 19 01 55.4 & $-04$ 29 12 & 10.966 & 0.007 & 10.658 & 0.008 & 10.575 & 0.008\\
S234-E & 20 31 20.4 & $-49$ 38 58 & 12.464 & 0.011 & 12.127 & 0.008 & 12.070 & 0.007\\
S813-D & 20 41 05.1 & $-05$ 03 43 & 11.479 & 0.005 & 11.142 & 0.005 & 11.085 & 0.005\\
P576-F & 20 52 47.3 & $ 06$ 40 05 & 12.247 & 0.004 & 11.940 & 0.004 & 11.880 & 0.005\\
S889-E & 22 02 05.7 & $-01$ 06 02 & 12.021 & 0.005 & 11.662 & 0.004 & 11.585 & 0.005\\
S677-D & 23 23 34.4 & $-15$ 21 07 & 11.857 & 0.003 & 11.596 & 0.003 & 11.542 & 0.003\\
P290-D & 23 30 33.4 & $ 38$ 18 57 & 11.634 & 0.005 & 11.337 & 0.004 & 11.262 & 0.006\\
\hline
         \end{tabular}
         \end{center}
\end{table*}

\begin{acknowledgements}\scriptsize
The authors acknowledge the reviewer for the useful comments. The authors also acknowledge Poshak Gandhi of Institute of Astronomy (University of Cambridge) for the useful codes used in the preliminary part of the study and Marc Sarazin of ESO for the useful comments on the GOES satellite data.
\end{acknowledgements}

\end{document}